%% file: main_arxiv.tex
\documentclass[12pt]{article}
\usepackage{graphicx} 
\usepackage{amsmath}
\usepackage[margin=1in]{geometry}
\usepackage{tikz}
\usepackage{setspace}
\usepackage{colortbl} 
\usepackage{xcolor}   

\usepackage{soul}

\usepackage{authblk} 

\usepackage{hyperref}
\usepackage{verbatim, color,
amssymb, float, epsfig}
\usepackage{caption}
\usepackage{subcaption}
\usepackage[style=apa,doi=false,url=false,hyperref=true,natbib]{biblatex}
\usepackage{xr}
\usepackage{cleveref}

\newcommand{\ba}  {\begin{array}}
\newcommand{\ea}  {\end{array}}
\newcommand{\be}  {\begin{equation}}
\newcommand{\ee}  {\end{equation}}
\newcommand{\bea}  {\begin{eqnarray}}
\newcommand{\eea}  {\end{eqnarray}}
 
\newcommand{\utheta}      {\mbox{\boldmath$\theta$}}

\newcommand{\uiota}       {\mbox{\boldmath$\uiota$}}

\newcommand{\unu}        {\mbox{\boldmath$\nu$}}

\newcommand{\utau}       {\mbox{\boldmath$\tau$}}

\def\boxit#1{\vbox{\hrule\hbox{\vrule\kern6pt
 \vbox{\kern6pt#1\kern6pt}\kern6pt\vrule}\hrule}}
\title{Evaluating Bias Reduction Methods in Binary Emax Model for Reliable Dose-Response Estimation}
\author{Jiangshan Zhang$^{1}$,  Vivek Pradhan$^{2}$, Yuxi Zhao$^{2}$  \\
        \small $^{1}$Department of Statistics, University of California, Davis \\
        \small $^{2}$I\&I and Specialty Care Statistics, Pfizer Inc.,
        \small 1 Portland Street, Cambridge, MA 02139, USA }

\providecommand{\keywords}[1]
{
  \small	
  \textbf{\textit{Keywords---}} #1
}

\date{}

\begin{document}

\maketitle

\begin{abstract}
  The Binary Emax model is widely employed in dose-response analysis during Phase II clinical studies to identify the optimal dose for subsequence confirmatory trials. The parameter estimation and inference heavily rely on the asymptotic properties of Maximum Likelihood (ML) estimators; however, this approach may be questionable under small or moderate sample sizes and is not robust to violation of model assumption. To provide a reliable solution, this paper examines three bias-reduction methods: the Cox-Snell bias correction, Firth’s score modification, and a maximum penalized likelihood estimator (MPLE) using Jeffreys prior. Through comprehensive simulation studies, we evaluate the performance of these methods in reducing bias and controlling variance, especially when model assumptions are violated. The results demonstrate that both Firth’s and MPLE methods provide robust estimates, with MPLE outperforming in terms of stability and lower variance. We further illustrate the practical application of these methods using data from the TURANDOT \citep{vermeire2017anti} study, a Phase II clinical trial. Our findings suggest that MPLE with Jeffreys prior offers an effective and reliable alternative to Firth’s method, particularly for dose-response relationships that deviate from monotonicity, making it valuable for robust parameter estimation in dose-ranging studies.
\end{abstract}


\section{Introduction}
In drug development, accurately estimating the dose-response relationship for an experimental compound is a critical step. This process involves assessing how varying doses of the compound influence its therapeutic effects, enabling researchers to identify the optimal dosage that maximizes efficacy while minimizing risks or adverse effects. Phase II clinical trials, particularly early-stage multigroup parallel trials, are designed to evaluate the efficacy of a compound across a range of dose levels. In analyses, the Emax model is a popular choice to characterize non-linear dose-response relationships. It can be implemented for various efficacy endpoints approved by the regulatory agency, such as binary (e.g., responder vs. non-responder), continuous (e.g., biomarker change from baseline), and time-to-event (e.g., progression-free survival, overall survival).  In this paper we focus on a dichotomized endpoint, which makes the binary Emax model an appropriate choice for analysis. Extensions to other endpoint types follow analogous likelihood-based formulations and are left to future work.

Let $n$ denote the total sample size, and suppose $y_i$ denotes the binary outcome for the $i$-th patient randomized to a dose $Dose_i$, where $Dose_i$ is one of the predefined dose levels from a set of $M$ levels $\{D_1,\cdots,D_M\}$, for $i=1,\cdots,n$. Without loss of generality, let $P(y_i =1|Dose_{i})= \pi_{i}$ be the probability of success for patient $i$ after dosage. Without other covariates, $\pi_i$ would be one of $M$ possible probability levels, corresponding to administered dose levels. Under this setup, the sigmoid four-parameter Emax model can be written as the following 
\be
log\left(\frac{\pi_{i}}{1-\pi_{i}}\right)=E_0+\frac{E_{max}\times Dose_{i}^{\lambda} }{ED_{50}^{\lambda} +Dose_{i}^{\lambda} }
\ee  
where $E_0$ is the expected logit of dose effect at $Dose_{i}=0$, with $Dose_{i}=0$ often being considered as placebo; $E_{max}$ is the expected logit of the maximum achievable effect (at infinite dose, can be positive or negative); $ED_{50}$ is the dose that produces half-maximal effect $E_{max}/2$  (typcially $ED_{50}>0$); and $\lambda$ is the positive slope factor or Hill parameter, determining the steepness of the dose-response curve: larger $\lambda$ yields a sharper,
more sigmoidal transition around $ED_{50}$. Despite its added flexibility, the four-parameter Emax model is often unnecessary; meta-analyses report that the three-parameter Emax model with $\lambda=1$ fits most empirical dose–response datasets comparably well \citep{Thomas20172,Kirby2011,Wu2017}. Hence the Emax model is reduced to
\be
\log\left(\frac{\pi_{i}}{1-\pi_{i}}\right)=E_0+\frac{E_{max}\times Dose_{i} }{ED_{50}+Dose_{i} }.
\label{eq:3param}
\ee

While the maximum likelihood estimates (MLE) of the three parameters are asymptotically consistent under regulatory conditions, it is well known that the estimates tend to be biased with finite sample size. For the binary Emax model,  the estimation process is more complicated and requires careful attention to several key issues: 

\emph{1. Separation in binary models.}When fitting binary outcome models, a common issue that may arise in a small sample setup is the non-convergence of estimates. This phenomenon is often referred as “separation” (\cite{ALBERT1984}). Complete separation occurs when a single covariate or a linear combination of covariates perfectly predicts the outcome, leading to divergence in the estimation process. Even in the absence of complete separation, quasi-complete separation—where a subset of subjects’ responses is perfectly predicted—can still pose significant challenges for estimation (\cite{altman2004numerical}). Practical warning signs of separation are: non-convergnece of the estimation algorithm, very large estimated coefficient and standard error pairs, and fitted probabilities numerically equal to 0 or 1 for some pattern of predictors. Separation can be diagnosed formally via linear-programming tests \citep{ALBERT1984} or with modern implementations that detect separation before fitting (e.g., \texttt{detect\_separation} function in R package \texttt{brglm2}; see \citealp{Kosmidis2020}). Although separation is common in biomedical research data, it is often overlooked, particularly when sample sizes are limited. For instance, existing methods and packages that utilize maximum likelihood estimation for dose-response or Emax models, such as the \textit{ClinDR} package in R developed by \citet{Thomas2017} and the \textit{Dosefinding} package by \citet{Bornkamp2010}, do not address the issue of separation.

\emph{2.Non-monotonic/Convex increasing sample response curve.} It is known that the Emax model is a simple concave function that increases monotonically with dose. However, sampling variability from finite small sizes can also cause maximum likelihood estimation non-convergence. This issue is well-documented for continuous outcomes, and several studies have examined the convergence problems of estimation algorithms (\cite{Flournoy2020, Flournoy2021, Chen2023}). It has been studied theoretically that the maximum likelihood method tends to fail in two scenarios: when the sample data follow a non-increasing concave shape, or when the data follow a convex increasing shape (\cite{Aletti2024}). In a practical view, when this issue arises, one can always provide a sensitivity fit with a more flexible candidate, such as a quadratic-logit model or beta model. A detail demonstration is given in the supplementary material.

To address these two issues in the framework of the Emax model, modified likelihood estimation methods were often considered. \citet{heinze2002solution} and \citet{heinze2006comparative} demonstrated that Firth’s modified score equations method (\cite{FIRTH1993}), originally developed to reduce bias in maximum likelihood estimates (MLE), can also effectively handle the issue of separation within the framework of generalized linear models (GLMs). Although it has been investigated for the first ill-defined scenario of the Emax model mentioned above, Firth's method cannot provide admissible estimations for Emax parameters. Even though it can produce a finite estimate under the second scenario mentioned above (\cite{Aletti2024}), this score adjustment is not equivalent to maximizing a penalized likelihood in nonlinear models such as Emax. This complicates implementation for methods that require explicit likelihood functions.

It draws our attention that modifying the likelihood with Jeffreys invariant prior is equivalent to Firth's method in the GLM framework with a canonical link function. In specific, for GLMs with canonical links, Firth’s modified score function coincides with the first order derivative of Jeffreys-prior penalized log likelihood, yielding shrinkage toward the canonical center \citep{FIRTH1993}. However, this equivalence doesn't hold when the link function is not canonical or the predictor relationship is nonlinear. \citet{Kosmidis2020} showed that the Jeffreys-prior penalty yields finite estimates for binomial-response GLMs with non-canonical links, and shrinks fitted probabilities away from 0 and 1. This shrinkage reduces the finite-sample bias of the MLE.  Nevertheless, the performance of Jeffreys prior penalty in non-linear regression with binary outcome has not been investigated or compared to Firth's method. In this paper, we derive the analytical expression for Jeffreys prior penalty, together with Firth’s correction for the score function of the binary outcome Emax model with a logistic link. We compare and quantify both approaches, along with the Cox-Snell bias correction method, in terms of their stability and consistency when estimating model parameters under small sample sizes.

\emph{ Relation to recent penalized likelihood for dose–response.} Contemporary work uses penalization in two complementary ways. First, bias-reduction penalties deliver finite, shrunken estimates in separation-prone binomial models and are now standard in practice \citep{Kosmidis2020}. Second, flexible dose–response modeling often proceeds via penalized splines (or related smoothers), which stabilize estimation by penalizing curvature and are widely used in dose–response and meta-analytic settings \citep{Kirby2009}. Our contribution fits within the first case, but in a nonlinear binary dose–response setting: we quantify how Jeffreys' penalization behaves under small samples and partial separation, and contrast it with Firth's modification and Cox-Snell correction.

The remainder of the paper is organized as follows: Section 2 outlines the approaches for bias reduction. Section 3 explores the simulation settings and presents the results. Section 4 discusses a real-world application, providing insights into the practical implementation of the bias reduction methods. Finally, Section 5 offers a discussion of the findings and concludes the paper. 

\section{Method}
For a consolidated list of symbols used throughout Sections 2–4, see Table S1 in supplementary material.
\subsection{Cox-Snell's bias correction}
Since the MLE can be biased in small-sample experimental settings, we therefore use the Cox–Snell expansion to obtain an analytic bias term, and subtract it from the MLE prior to reporting estimates and uncertainty.
With some regularity conditions of the likelihood functions, the MLE is estimated by solving the score function $U(\hat\theta)=0$. By Taylor expansion around the true $\utheta$, we can show that:
\[
\hat\utheta-\utheta \;=\; I(\utheta)^{-1}\Big\{\tfrac{1}{n}U(\utheta)\Big\}
\;-\;\tfrac{1}{2}\,I(\utheta)^{-1}\, \Big[\tfrac{1}{n}\, \partial^2 U(\utheta)/\partial\theta\partial\theta^\top \Big]\,
I(\utheta)^{-1}\,U(\utheta) \;+\; \cdots .
\] 
Taking expectations on both sides, the second term on the right produces a nonzero contribution of order $\mathcal{O}(n^{-1})$, which leads to the bias. We can rewrite the Bias of MLE of a vector of $p$ parameters
as $\mathrm{E}(\hat{\utheta}-\utheta)= \mathbb{B}(\utheta)+\mathcal{O}(n^{-2})$, where $\mathbb{B}(\utheta)$ is the epexcation of the second term.  \citet{Cox1968} proposed a general form of $\mathbb{B}(\utheta)$ with
\begin{equation}
    \mathbb{B}\left({\utheta}_s\right)=\sum_{r=1}^p \sum_{j=1}^p \sum_{l=1}^p \kappa^{s r} \kappa^{j l}\left[\frac{1}{2} \kappa_{r j l}+\kappa_{r j, l}\right]
\end{equation}
where ${\utheta}_s$ is the $s$th element of 
$\utheta$, $\kappa^{r j}$ is the $(r,j)$th element of the inverse of the negative expected Fisher information matrix, $\kappa_{r j l}$ and $\kappa_{r j, l}$ are joint cumulants of the derivatives of the log-likelihood $l$ expressed in the following. Denote $H_{rj}$ as the $(r,j)$th element of the Hessian matrix, and $U_l$ as the $l$th element of the score function vector:
\begin{equation}
    \begin{split}
        \kappa_{r j l} &= \mathrm{E}\left[\frac{\partial^3 l}{\partial \utheta_r \partial \utheta_j \partial \utheta_l}\right] = \mathrm{E}\left[\frac{\partial H_{rj}}{\partial \utheta_l}\right],\\
        \kappa_{r j, l} &= \mathrm{E}\left[\left(\frac{\partial^2 l}{\partial \utheta_r \partial \utheta_j}\right)\left(\frac{\partial l}{\partial \utheta_l}\right)\right] =  \mathrm{E}\left[ H_{rj} U_{l}\right].
    \end{split}
\end{equation}
By substituting $\utheta$ with the MLE $\hat{\utheta}$, the Bias corrected MLE can be expressed as: 
\begin{equation}
    \hat{\utheta}_c=\hat{\utheta}-\mathbb{B}(\hat{\utheta}).
\end{equation}
Under the setting of the logit Emax model, the $p$-dimensional score function vector is expressed as
\begin{equation}
    U(\boldsymbol{\theta})=\sum_{i=1}^n \left(y_i-\pi_{i}\right) \nabla \eta\left(\text {Dose}_{i}, \boldsymbol{\theta}\right)
\end{equation}
with $\utheta=(E_0, E_{max}, ED_{50})$, and $\nabla \eta\left(\text {Dose}_{i}, \boldsymbol{\theta}\right)$ given by
$$
\nabla \eta\left(\text {Dose}_{i}, \boldsymbol{\theta}\right)=\left(1, \frac{\text {Dose}_{i}}{\text {Dose}_{i}+E D_{50}},-\frac{\text {Dose}_{i} \times E_{\max }}{(\text {Dose}_{i}+E D_{50})^2}\right)^T.
$$
Accordingly, the $p \times p$ Hessian matrix $H(\utheta)$ is formulated as follows:
\begin{equation}
    H(\boldsymbol{\theta})=\sum_{i=1}^n \left(\pi_{i}-1\right) \pi_{i} \nabla \eta\left(\text {Dose}_{i}, \boldsymbol{\theta}\right)^{\top} \nabla \eta\left(\text {Dose}_{i}, \boldsymbol{\theta}\right)-A_i(\boldsymbol{\theta})
\end{equation}
where
$$
A_i(\boldsymbol{\theta})=\left(\begin{array}{ccc}
0 & 0 & 0 \\
0 & 0 & \frac{\left(y_i-\pi_{i}\right) \text {Dose}_{i}}{\left(ED_{50}+\text {Dose}_{i}\right)^2} \\
0 & \frac{\left(y_i-\pi_{i}\right) \text {Dose}_{i}}{\left(ED_{50}+\text {Dose}_{i}\right)^2} & -\frac{2\left(y_i-\pi_{i}\right) \text {Dose}_{i} \times E_{\max}}{\left(ED_{50}+\text {Dose}_{i}\right)^3}
\end{array}\right),
$$
and the expected Fisher information matrix $I(\utheta)$ is represented as follows:
\begin{equation}
    I(\utheta) = \sum_{i=1}^n \left(1-\pi_{i}\right) \pi_{i} \nabla \eta\left(\text {Dose}_{i}, \boldsymbol{\theta}\right)^{\top} \nabla \eta\left(\text {Dose}_{i}, \boldsymbol{\theta}\right).
    \label{eq::expected_inf}
\end{equation}
The closed-form derivations of $\kappa_{r j l}$ and $\kappa_{r j, l}$ of the logit Emax model are provided in the supplementary material. In the general case for high-dimensional parameters, when closed forms are tedious, we can use automatic differentiation with Vector Jacobian products to evaluate directional third derivatives. See \citet{GriewankWalther2008} for details.

It is important to note that the Cox-Snell approach is a post-hoc correction approach, which first requires the MLE to be estimated, followed by the computation of the bias term. This method fails if the iterative algorithm for calculating the MLE does not converge, such as in cases with binary response data that exhibit complete separation. Another issue arises when the observed dose-response relationship does not display a strongly concave increasing pattern. When the curve is nearly flat over the observed doses, the eigenvalues of the Fisher information are small, or even the information matrix is ill-conditioned. This issue with the information matrix causes divergence of the algorithm. Even if the algorithm converges, quasi-separation in the dataset can result in large variations in MLE estimation, causing the bias reduction method to perform poorly. This is because the bias correction partly depends on the inverse of the Fisher information, which also becomes unstable in quasi-separation scenarios.

\subsection{Firth's score modification}
Instead of applying a post-hoc bias correction to the MLE, Firth proposed a preventive approach that directly estimates a bias-controlled estimator, helping to avoid issues of non-convergence and large variance in estimates \citep{FIRTH1993}. This method directly modifies the score function by adding an adjustment term. Denote the modified score function $\tilde{U}_s$, original score function $U_s$ and the addition term $W_s$. The expression of $\tilde{U}_s$ is given by:
\begin{equation}
\tilde{U}_s=U_s+W_s, \quad s=1, \ldots, p.
\end{equation}
The addition term $W_s$ is expressed as below:
$$W_s=\frac{1}{2} \operatorname{tr}\left\{I^{-1}\left(P_s+\kappa_{..,s}\right)\right\}$$
with $P_s=\mathrm{E}\left(UU^{\top} U_s \right)$ and $\kappa_{..,s}=\mathrm{E}\left(H U_s \right)$.
It can be seen that $P_s$ is related to the third central moment of the Bernoulli variable, which is $\mathrm{E}\left[(y_i-\pi_{i})^3\right]=(1-\pi_{i})^3\pi_{i}-(1-\pi_{i})\pi_{i}^3$. This introduces complexity in the binary outcome models as compared to models with normal distributions. Under normality, the third moment is always zero, which means $P_s$ is not included in the modification term for regression models with normal errors (\cite{Aletti2024}). In the framework of the logit Emax model, as described by \citet{McCullagh1989}, we have 
\begin{equation}
    P_s = \sum_{i=1}^n \frac{\kappa^{(3)}_{i}}{\kappa^{(2)}_{i}}I [\nabla \eta\left(\text {Dose}_{i}, \boldsymbol{\theta}\right)]_s = \sum_{i=1}^n(1-2\pi_{i})I [\nabla \eta\left(\text {Dose}_i, \boldsymbol{\theta}\right)]_s;
\end{equation}
where $\kappa^{(k)}_{i}$ is the $k$th cumulant of the ith response $y_i$, which are the same as central moments of $y_i$ in the second and third order. Regarding $\kappa_{..,s}$, as the first cumulant of response $\mathrm{E}\left[(y_i-\pi_{i})\right]=0$, under the assumption of independence between observations, it can be shown that
\begin{equation}
    \kappa_{..,s} = \sum_{i=1}^n\mathrm{E} \left(-A_i(\boldsymbol{\theta}) U_{is}\right) = \sum_{i=1}^n (1-\pi_{i})\pi_{i} Q_i(\utheta) [\nabla \eta\left(\text {Dose}_{i}, \boldsymbol{\theta}\right)]_s
\end{equation}
where 
$$
Q_i(\boldsymbol{\theta})=\left(\begin{array}{ccc}
0 & 0 & 0 \\
0 & 0 & \frac{ \text {Dose}_{i}}{\left(ED_{50}+\text {Dose}_{i}\right)^2} \\
0 & \frac{ \text {Dose}_{i}}{\left(ED_{50}+\text {Dose}_{i}\right)^2} & -\frac{2 \text {Dose}_{i} \times E_{\max}}{\left(ED_{50}+\text {Dose}_{i}\right)^3}
\end{array}\right).
$$
With the score modification, a bias-reduced estimator can be obtained using linear equation-solving algorithms, such as Newton and Broyden. It is worth noting that Firth's modification here applies to the score function, and because the derivative matrix of $\tilde{U}_s$ is not symmetric, there is no penalized likelihood corresponding to the modified score. As a result, optimization algorithms such as Newton-Raphson can not be implemented.

\subsection{Maximum penalized likelihood estimation with Jeffreys prior}
While the modification of the score function can yield a bias-reduced estimator, the lack of an explicit form for the modified likelihood function limits the method's applicability. For instance, the expectation-maximization method for the mixture model can not be implemented with the modified score function. \citet{FIRTH1993} demonstrated that for the canonical parameter in the full exponential family and generalized linear models, the score function modification approach is equivalent to estimation by maximizing the penalized likelihood function using the Jeffreys prior as the penalty term (referred to as Maximum Penalized Likelihood Estimator or MPLE). However, these assumptions do not hold for the curved exponential family or the nonlinear regression settings, such as the logit Emax model. Therefore, it is of interest to assess the bias control performance of MPLE in these underexplored contexts. \citet{Kosmidis2020} proved the finiteness and the shrinkage of Jeffreys prior for curved exponential families with the linear regression setting, which formally validated the bias-control of the MPLE. However, its performance in nonlinear regression models remains unclear. The rest of the section outlines the details and offers a more in-depth discussion in the context of logit Emax model.

The penalized likelihood function for the Emax model is obtained by multiplying the likelihood function by Jeffreys invariant prior as a penalty term, which can be expressed as:
\begin{equation}
    L^*(\utheta|y,Dose) = L(\utheta|y,Dose)|I(\utheta)|^{1/2}.
    \label{pll:y}
\end{equation}
Note that the score function for the penalized likelihood differs from the Firth modified score function. It can be seen that the Jeffreys prior, which is the square root of the determinant of the information matrix, is always finite and positive, thereby reducing the variability of the estimators. The logarithm of the likelihood and the score function of the penalized log-likelihood with respect to $\utheta_s$ is given by:
\begin{equation}
     U_s^*(\utheta) = \sum_{i=1}^n (y_i-\pi_{i}) \nabla \eta(\text {Dose}_{i},\utheta) + \mathrm{tr}\left(I^{-1} \frac{\partial I}{\partial \utheta_s} \right ).
\end{equation}
As can be seen, the difference between ${\partial I}/{\partial \utheta_s}$ presented above and $P_s+\kappa_{.,s}$ given in Firth's score modification distinguishes the two methods, and it can be derived that
\begin{equation}
  {\partial I}/{\partial \utheta_s}-  P_s-\kappa_{.,s} = \kappa_{.,s}.
\end{equation}
While $U_s^*(\utheta)$ remains biased in $\mathcal{O}(n^{-1})$ in general, the constant term of the bias becomes significantly smaller, making the estimation more reliable. In fact, \citet{Box1971} demonstrated that for non-linear regression with normal errors, the bias term $\mathbb{B}$ is approximately equal to $1/2 I^{-1} tr(I^{-1} \partial I/\partial \theta)$, which, when converted back to the likelihood modification, is exactly the Jeffreys prior. Studies by \citet{Box1971}, \citet{Bates1980}, \citet{Clarke1980}, \citet{Hougaard1985} and \citet{Amari1982} have shown that controlling the bias term results in an actual bias that closely matches the specific bias approximation obtained through simulation. Therefore, without considering the additional bias term $\kappa_{\cdot,s}$ introduced by the non-linearity under the generalized model setting, in practice, Jeffreys prior penalization performs as well as Firth's score modification.

Hereafter, we denote the MLE as $\hat{\utheta}$, the Cox-Snell estimator as $\hat{\utheta}_{C}$, the Firth's score modified estimator as $\hat{\utheta}_{F}$, and MPLE with Jeffreys prior as $\hat{\utheta}_{J}$.
\subsection{Variance estimator and confidence interval}
Using the asymptotic properties of the MLE, the estimated variance-covariance matrix can be derived from the observed information matrix evaluated at the estimated MLE $\hat{\theta}$ in Equation \eqref{eq::expected_inf}, as shown below:
\begin{equation}
    \widehat{Var}(\hat{\utheta}) = \left[-H(\hat{\utheta})\right]^{-1}.
    \label{eq::var}
\end{equation}
Since neither the Cox-Snell nor the Firth's score modified estimation approaches alter the original likelihood function, the variance-covariance matrices for these two estimators, namely $\hat{\utheta}_{C}$ and $\hat{\utheta}_{F}$, can be still derived using the observed information matrix evaluated at their respective estimates by Equation \eqref{eq::var}. For Jeffreys penalized likelihood maximization approach, the observed information matrix based on the penalized log-likelihood is given as,
\begin{equation}
   O^*(\utheta) = -H^*(\utheta) = -\frac{\partial U^*(\utheta)}{\partial \utheta}.
\end{equation}
Consequently, a consistent estimator of variance-covariance
matrix of the $\hat{\utheta}_{J}$ is then obtained as the inverse of $O^*(\hat{\utheta}_{J})$.

By utilizing the variance-covariance matrix estimator of each estimator mentioned above, namely $\hat{\utheta}$, $\hat{\utheta}_{C}$, $\hat{\utheta}_{F}$, and $\hat{\utheta}_{J}$, confidence intervals for the $\utheta$ can be constructed accordingly. The standard errors of parameters in $\utheta$ are calculated by taking the square root of the corresponding diagonal elements of the variance-covariance matrix estimator, and the 100(1-$\alpha$)\% confidence interval based on the normal approximation for parameter $\utheta_s$ is shown as below:

\begin{equation}
    \left(\hat{\utheta}_s -z_{\alpha/2}\sqrt{\widehat{V}_{s,s}},\hat{\utheta}_s +z_{\alpha/2}\sqrt{\widehat{V}_{s,s}}\right)
    \label{eq::ci}
\end{equation}
where $z_{\alpha/2}$ represents the $1-\alpha/2$ quantile of the standard normal distribution and $\widehat{V}$ denotes the variance-covariance matrix estimator.

\section{Simulation study}
To evaluate the performances of the three bias-reduction estimation methodologies, simulation studies were designed and conducted. The simulation was based on a general phase-2 dose-response clinical trial setting. To investigate the effect of sample size on the estimations, we considered four sample sizes: $n = 50,100, 150$, and 200. The sample size budget was evenly allocated across five different treatment dose arms: Dose=$(0, 7.5, 22.5, 75, 225)$, ensuring equal sample sizes in each arm. We set the success rate for response in the placebo arm ($\text {Dose}_{i}=D_1=$0) to be 10\%, the maximum success rate at the infinite dosage was set to be 80\%, and the dosage which produces a half-maximal effect was set to 7.5. In other words, we set $E_0=\mathrm{logit}(0.1)=-2.197$, $E_{max}=\mathrm{logit}(0.8)-\mathrm{logit}(0.1)=3.583$, and $ED_{50}=7.5$ (log($ED_{50}$)=2.015).
The response variable $y_i$ was generated from a Bernoulli distribution with success probability $\pi_{ij}$, following the three-parameter logistic Emax model, for $i=1,...,n$. To enforce positivity estimation of $ED_{50}$, we fit the model using the log-parameterization for all methods. The starting value for all methods was set as the same, using {\it startemax} function in R package {\it ClinDR}. Optimization algorithms were employed with the same stopping rules for all methods:
gradient of objective $\le 10^{-6}$ or relative parameter/objective change $\le 10^{-8}$,
with a maximum 2{,}000 iterations. For each sample size scenario, $N=1000$ simulation replications were performed.

Table \ref{Sim:fail_proportion} shows the proportion of occurrences of non-convergence in log-likelihood estimation, along with the proportion of unstable estimations due to large variance across all four methods. Here, we define unstable estimation if any of the following hold: (i) $\hat{ED}_{50}$ hits prespecified bounds (i.e. $\hat{ED}_{50}>10\,D_{\max}$ or or $< 0.02\,D_{\min}$) and (ii) standard error of any parameter is undefined/NA, or relative standard error $\mathrm{SE}(\hat\utheta_s)/(|\hat\utheta_s|) > 5$. In the subsequent discussion, we examined the results of the sample size $n =50$, where separation occurs more frequently and the chance of observing a non-concave increasing dose-response relationship is not negligible.  As mentioned earlier, non-convergence can result from complete separation or non-concave increasing patterns in the dose-response data, while unstable estimation can arise from quasi-separation or slight violations of the concave increasing assumption. For both the MLE and Cox-Snell methods, 19.3\% of the simulations fail to yield estimates. Additionally, in 15.2\% of cases, while MLE could be obtained, it resulted in ill-posed fisher information, leading to unstable bias correction in the Cox-Snell estimates. Neither the Firth modification method nor the MPLE produced non-existent estimates; however, about 3\% of the Firth estimates were unstable due to the Fisher information issues, and only 2 replications using MPLE resulted in unstable penalized likelihood estimates. As the sample size increased, the frequency of estimation failures and unstable estimates decreased, although MLE and Cox-Snell correction still showed some instability. In contrast, both Firth’s score modification and MPLE consistently produced stable estimates when the sample size exceeded 150.

Table \ref{Sim:estimate} presents the estimation results for the four methods across four different sample size scenarios. For replications where methods failed to produce an estimate,  the value was recorded as NA and excluded from the metric calculations. For point estimation, we reported the average estimated value of $\utheta_s$, denoted as $\hat{\utheta_s} = 1/N\sum \hat{\utheta_s}^{(t)}$ for $s=1,2,3$, where $\hat{\utheta}_s^{(t)}$ is the estimates of $\utheta_s$ in $t_{th}$ replication. Similarly, we reported the mean bias error $\mathrm{MBE}=1/N\sum \hat{\utheta_s}^{(t)}-\utheta_s$, the mean squared error $\mathrm{MSE}=1/N\sum (\hat{\utheta_s}^{(t)}-\utheta_s)^2$, and the mean estimated standard error for $\hat{\theta}_i$, denoted as $\hat{\sigma_i} =1/N\sum \hat{\sigma_i}^{(t)}$. For confidence interval estimation, we reported coverage probability (CP) of 95\% confidence interval defined as $=1/N\sum I(\hat{\utheta_s}^{(t)})$, where $I(\cdot)$ is the indicator function for whether $\utheta_s$ falls within the estimated confidence interval. We also reported the mean estimated interval length (Est.length) computed as $=1/N\sum \lambda(\hat{\utheta_s}^{(t)})$, where $\lambda(\hat{\utheta_s}^{(t)})$ represents for the length of the confidence interval for $\hat{\utheta_s}^{(t)}$. 

As shown in Table \ref{Sim:estimate}, the Cox-Snell method produces highly unstable estimates when the sample size is small. Although the estimates become more stable as the sample size increases, extreme estimates still occur, impacting the mean metric performances. For MLE, among the replications where it converges, point estimates are stable but with a large bias in terms of MBE. Additionally, the variance estimates are extremely unstable for small sample sizes, which explains the poor performance of the Cox-Snell method. Although the variance estimates stabilize with larger sample sizes, they remain larger compared to Firth and MPLE. Both Firth and MPLE estimates show considerable stability even with small sample sizes, and both exhibit small biases across all sample size scenarios. Overall, MPLE consistently outperforms the other methods when evaluating MSE,  yielding the smallest standard errors across all scenarios. Furthermore, the 95\% confidence intervals constructed from these estimates show better coverage probability. As with many penalized likelihood estimators, the penalty term slightly reduces accuracy but significantly decreases variance, resulting in better performance in terms of mean squared error.

We also investigated how the varying values of parameters of the Emax model influence the estimation performances of the four methods under a moderate sample size. We fixed the sample size at $n = 200$ for all simulation scenarios to ensure relatively stable estimation. We varied the true maximal achievable effect to be 30\% ($E_{max}=1.349$), 50\%($E_{max}=2.197$), and 70\%($E_{max}=3.044$), while keeping the placebo effect fixed at 10\% and the half-maximal effect dose fixed at 7.5. As shown in Table \ref{Sim:estimate_emax}, all four methods perform better, with reduced bias and MSE as the maximal achievable effect increases. When the true maximal achievable effect is small, all four methods exhibit relatively large bias in estimating $ED_{50}$, compared to the other two parameters. When evaluating MSE, the Cox-Snell estimator sometimes has a higher MSE compared to MLE due to unstable variance estimates. In contrast, MPLE has a stable lower MSE compared to Firth's estimator, especially when the maximal achievable effect is small. Although the estimated standard error of MPLE is the smallest among all four estimators, it slightly overestimates the standard error when the maximal achievable effect is 30\%.

Next, we altered the half-maximal effect dosage to be 25, 50, and 75, with the placebo effect fixed at 10\%, and the maximal achievable effect to be fixed at 50\%. Table \ref{Sim:estimate_ed50} illustrates that as the true half-maximal effect dosage increases, the biases for all four methods also rise, with the Cox-Snell estimator exhibiting the largest bias even when compared to MLE. While Firth's estimator provides bias-reduced estimates of $E_{max}$, its performance is inferior to that of MLE when estimating $ED_{50}$ and $E_0$. Although the strength of bias reduction is modest, MPLE with Jeffreys prior still shows lower bias compared to MLE, particularly for $E_{max}$. In terms of MSE, MPLE consistently exhibits the smallest value among all estimators. In contrast, both the Cox-Snell estimator and Firth's estimator show no improvement over MLE, and in some scenarios, they even produce higher MSE. This underscores the instability of these two methods in the estimation of higher-order moments with a potential non-monotone concave increasing pattern of dose-response, particularly due to the small sample size.

Additional simulation results, including sample response shape-stratified performance (convex increasing and non-monotonic increasing), are provided in the Supplementary Material.

\section{Real data analysis}
To illustrate the application of the aforementioned bias reduction methods in a practical context, we analyzed data from the TURANDOT study (\cite{TURANDOT}), a Phase II randomized, double-blind, placebo-controlled clinical trial targeting ulcerative colitis in patients with moderate to severe symptoms. In this trial, 357 patients were randomly allocated to either a placebo group or one of four active dose groups, namely 7.5 mg, 22.5 mg, 75 mg, and 225 mg. As reported by \citet{TURANDOT}, the study exhibited a non-monotonic dose-response curve, where the highest dose group (225 mg) demonstrated reduced efficacy. Overall, the actual number of patients randomized into each of these treatment groups is approximately evenly allocated. The primary endpoint was clinical remission at week 12, with some instances of missing data, as detailed in Table \ref{Table::turandot_descriptive}.

For our analysis, we employed the Emax model, excluding patients with missing remission data, thereby performing a complete case analysis. We compared the four estimation methods and the results are shown in Table \ref{Table::turandot_analysis}. All four estimation methods converged, yielding point estimates. However, MLE, Cox-Snell estimator, and Firth's estimator show unstable estimation in terms of standard error when estimating $log(ED_{50})$. Additionally, the estimation of $E_{max}$ using the Cox-Snell estimator exhibited greater variation compared to the other methods. This instability may be attributed to the non-concave increasing pattern observed in the dose-response data. Notably, even when the 225mg dose arm with reduced efficacy was excluded from the analysis, the 75mg dose arm still displayed a slightly reduced efficacy pattern. This resulted in a non-increasing concave dose-response shape, leading to unstable variance estimates for the MLE and inconsistent expectations for the score function. On the other hand, MPLE, which incorporates a penalization term based on expected information, controls the variance of the estimator in a similar way as other penalized methods, for example, ridge regression. As a result, both point estimates and variance estimates from this method are relatively stable. 

\begin{figure}[htbp]
    \centering
    \includegraphics[width=1\linewidth]{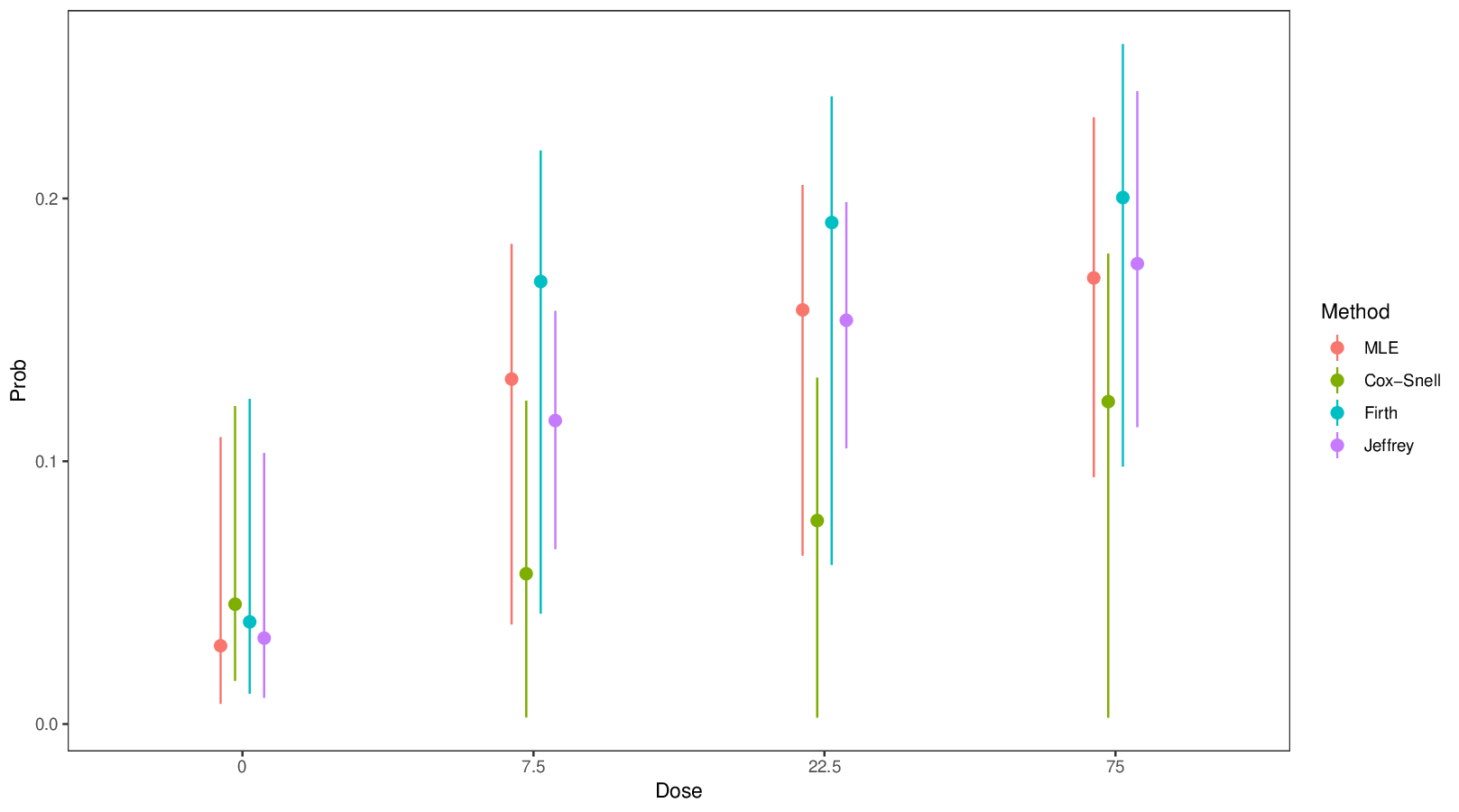}
    \caption{Estimated dose response remission probabilities based on different methods with their bootstrapped 95\% confidence intervals.}
    \label{fig:dose_response}
\end{figure}

To further evaluate the performance of these methods, we estimated the probabilities of remission and calculated their bootstrapped 95\% confidence intervals using 5000 bootstrap samples. As shown in Figure \ref{fig:dose_response}, despite the Cox-Snell method yielding the highest estimated mean probability for the placebo group, its estimated probabilities for the other dose groups were consistently lower than those from the other three methods. In contrast, Firth's score-modified approach yields the highest estimated mean probabilities across the three active dose groups, while the estimated probability for the placebo group was similar to those from MLE and MPLE. Both the Cox-Snell and Firth's score modification approaches exhibit significant variability in their estimates, particularly at higher dose levels. MPLE approach produced estimated mean probabilities close to those from MLE, but with much smaller variance, especially in the 7.5 mg and 22.5 mg dose groups.

\section{Conclusion and discussion}
The separation issues induced by the binary response model and unstable convergence of Emax model due to the non-monotonic increase of the dose-response relationship raise difficulties in the estimation procedure under a small or moderate sample size. In this paper, we addressed the challenge of robustly estimating the coefficients of the binary Emax model in the context of a small or moderate sample size by deriving a bias-corrected point estimator for the Emax model parameters utilizing the Cox-Snell approach, a bias-reduced point estimator using Firth’s score modification, and a penalized likelihood estimator using Jeffreys prior(referred as MPLE). Our simulation studies show that the two bias-preventive methods—Firth’s score-modified approach and the proposed MPLE with Jeffreys prior—consistently reduced bias across all sample sizes. Furthermore, MPLE exhibited less variation than Firth’s score modification in cases where the assumptions of the Emax model were slightly violated. Overall, point estimators from both bias-preventive approaches are robust against separation and minor violations of the monotone increasing pattern of the Emax model, in contrast to the instability and non-convergence observed with MLE and Cox-Snell estimation.


It is critical to notice that the potential observed non-monotonic increasing pattern of the dose-response relationship may not be due to the sampling uncertainty induced by small or moderate sample sizes, but the effect of the response missingness, especially when missingness is non-ignorable, as shown in \citep{TURANDOT}. In this paper we focus on separation and small-sample bias, but we note the following practical workflow for incomplete outcomes. First, we need to decide the missing mechanism on a primary set of assumptions. Under the missing at random (MAR) mechanism, apply multiple imputation with the binary Emax model and report MI-combined estimates. Under MNAR, conduct prespecified sensitivity analyses using selection or pattern-mixture formulations\citep{Emerson2018}, and consider model-based estimation via an expectation–maximization (EM) algorithm. Penalized likelihood can be combined with these procedures to stabilize estimation in separation-prone designs. For the binary Emax model specifically, our companion work develops penalized MNAR estimation and provides code for implementation \citep{zhang2024}. A full integration of these missing-data strategies into the present framework is a valuable direction for future work.

Generally, approximate confidence intervals for the MPLE can be constructed by the usual Wald method. However, relying on symmetric Wald CIs from PLEs can be unwise, since the small samples and separation-prone designs that motivate PLEs typically yield nonquadratic log-likelihoods. The resulting Wald intervals are therefore slightly wider, and empirical coverage often exceeds the nominal level (see also \citealp{heinze2002solution, Bull2006}). Although conservative MPLE confidence intervals can act as a guardrail against over-optimistic efficacy claims, intervals based on the profile penalized likelihood can be a choice for more precise estimation. 

{\bf Acknowledgements.} The authors are grateful to an Associate Editor and referees, whose insightful comments have helped improve the work and presentation. The R package \textit{jpemax} is accessible at https://github.com/Celaeno1017/jpemax.

\printbibliography

\newpage

\begin{table}[htp]
    \centering
    \begin{tabular}{ll|rr}
    \hline
         Sample Size(n)&Type& Fail to estimate (\%) & Unstable estimate (\%) \\
         \hline
        50 & MLE & 19.3 & 15.2 \\
        & Cox-Snell & 19.3 & 15.2\\
        & Firth & 0 & 3.3\\
        & MPLE & 0& 0.2 \\
        100 & MLE & 3.8 & 4.2 \\
        & Cox-Snell & 3.8 & 4.2\\
        & Firth & 0 & 0.6\\
        & MPLE & 0& 0 \\ 
         150 & MLE & 0.8 & 1.0 \\
        & Cox-Snell & 0.8 & 1.0\\
        & Firth & 0 & 0\\
        & MPLE & 0& 0 \\ 
         200 & MLE & 0.6 & 0.4 \\
        & Cox-Snell & 0.6 & 0.4\\
        & Firth & 0 & 0\\
        & MPLE & 0& 0 \\ 
         \hline
    \end{tabular}
    \caption{Proportions of times that estimates do not exist and estimates are not stable with extreme values or variance across four methods.}
    \label{Sim:fail_proportion}
\end{table}

\begin{table}[ht]
\centering
\resizebox*{!}{0.9\textheight}{
\begin{tabular}{lll|rrrrrr}
  \hline
 Sample Size(n)&  Parameter &Type  & Estimate & MBE & MSE & Est.SE & CP & Est.Length \\ 
  \arrayrulecolor{black}\hline\arrayrulecolor[gray]{0.8}
50&log($ED_{50}$) & MLE  & 1.743 & -0.272 & 6.203 & 14.806 & 0.991 & 58.038 \\ 
   && Cox-Snell  & 63.463 & 61.449 & $5.116\times10^5$ & 22.323 & 0.835 & 87.506 \\ 
   && Firth  & 1.953 & -0.062 & 7.197 & 2.100 & 0.990 & 8.233 \\ 
   && MPLE  & 2.056 & 0.041 & 1.085 & 0.817 & 0.942 & 3.203 \\ 
   \hline
   &$E_{max}$& MLE  & 4.834 & 1.251 & 15.103 & 7.126 & 0.984 & 27.935 \\ 
   && Cox-Snell  & 70.337 & 66.753 & $2.904\times10^7$ & 5.156 & 0.781 & 20.211 \\ 
   && Firth  & 3.447 & -0.137 & 388.690 & 2.449 & 0.936 & 9.598 \\ 
   && MPLE  & 3.783 & 0.199 & 4.506 & 1.108 & 0.956 & 4.342 \\ 
   \hline
   &$E_0$& MLE  & -2.962 & -0.765 & 6.743 & 6.452 & 0.960 & 25.291 \\ 
   && Cox-Snell  & 934.812 & 937.009 & $1.376\times10^8$ & 5.169 & 0.826 & 20.223 \\ 
   && Firth  & -2.602 & -0.405 & 7.275 & 2.141 & 0.892 & 8.391 \\ 
   && MPLE  & -2.244 & -0.047 & 4.265 & 0.897 & 0.936 & 3.518 \\ 
   \arrayrulecolor{black}\hline\arrayrulecolor[gray]{0.8}
  100&log($ED_{50}$)& MLE  & 1.928 & -0.087 & 0.866 & 0.992 & 0.985 & 3.889 \\ 
   && Cox-Snell  & 6.430 & 4.415 & $1.364\times10^3$ & 0.864 & 0.947 & 3.386 \\ 
   && Firth  & 1.986 & -0.028 & 1.133 & 1.338 & 0.996 & 5.244 \\ 
   && MPLE  & 2.054 & 0.039 & 0.389 & 0.763 & 0.978 & 2.991 \\ 
   \hline
  &$E_{max}$ & MLE  & 4.186 & 0.602 & 3.735 & 2.159 & 0.972 & 8.465 \\ 
  && Cox-Snell  & -9.313 & -12.897 & $1.941\times10^4$ & 1.756 & 0.894 & 6.883 \\ 
   && Firth  & 3.763 & 0.180 & 1.089 & 0.990 & 0.970 & 3.881 \\ 
   && MPLE  & 3.679 & 0.096 & 0.673 & 0.888 & 0.958 & 3.482 \\ 
   \hline
  &$E_0$ & MLE  & -2.680 & -0.482 & 3.440 & 1.927 & 0.962 & 7.552 \\ 
   && Cox-Snell  & 19.947 & 22.144 & $6.863\times10^4$ & 1.646 & 0.896 & 6.452 \\ 
   && Firth  & -2.230 & -0.032 & 0.869 & 0.889 & 0.944 & 3.484 \\ 
   && MPLE  & -2.253 & -0.056 & 0.493 & 0.780 & 0.948 & 3.056 \\ 
   \arrayrulecolor{black}\hline\arrayrulecolor[gray]{0.8}
150&log($ED_{50}$) & MLE  & 1.957 & -0.058 & 0.458 & 0.649 & 0.982 & 2.546 \\ 
   && Cox-Snell  & 2.282 & 0.267 & 2.249 & 0.701 & 0.978 & 2.748 \\ 
   && Firth  & 2.043 & 0.028 & 0.554 & 0.716 & 0.976 & 2.808 \\ 
   && MPLE  & 2.058 & 0.043 & 0.296 & 0.609 & 0.978 & 2.388 \\ 
   \hline
  &$E_{max}$ & MLE  & 3.926 & 0.343 & 1.499 & 1.036 & 0.971 & 4.060 \\ 
   && Cox-Snell  & 2.708 & -0.875 & 50.335 & 0.946 & 0.941 & 3.708 \\ 
   && Firth  & 3.705 & 0.122 & 0.598 & 0.721 & 0.972 & 2.826 \\ 
   && MPLE  & 3.696 & 0.112 & 0.532 & 0.711 & 0.972 & 2.788 \\ 
   \hline
   &$E_0$ & MLE  & -2.460 & -0.262 & 1.294 & 0.888 & 0.969 & 3.480 \\ 
   && Cox-Snell  & -1.422 & 0.776 & 45.276 & 0.782 & 0.947 & 3.065 \\ 
   && Firth  & -2.217 & -0.020 & 0.449 & 0.639 & 0.962 & 2.507 \\ 
   && MPLE  & -2.272 & -0.074 & 0.387 & 0.629 & 0.968 &  2.464\\ 
     \arrayrulecolor{black}\hline\arrayrulecolor[gray]{0.8}
200&log($ED_{50}$) & MLE  & 1.971 & -0.044 & 0.305 & 0.534 & 0.966 & 2.094 \\ 
   && Cox-Snell  & 2.092 & 0.078 & 0.235 & 0.524 & 0.990 & 2.054 \\ 
   && Firth  & 2.037 & 0.022 & 0.350 & 0.547 & 0.964 & 2.146 \\ 
   && MPLE  & 2.044 & 0.029 & 0.225 & 0.520 & 0.964 & 2.039 \\ 
   \hline
  &$E_{max}$ & MLE  & 3.793 & 0.209 & 0.803 & 0.678 & 0.970 & 2.659 \\ 
   && Cox-Snell  & 3.486 & -0.098 & 0.710 & 0.648 & 0.958 & 2.540 \\ 
   && Firth  & 3.663 & 0.080 & 0.383 & 0.611 & 0.968 & 2.395 \\ 
   && MPLE  & 3.665 & 0.082 & 0.368 & 0.604 & 0.970 &2.367  \\ 
   \hline
  &$E_0$ & MLE  & -2.329 & -0.132 & 0.731 & 0.610 & 0.960 & 2.390 \\ 
   && Cox-Snell & -2.031 & 0.166 & 1.598 & 0.580 & 0.944 & 2.274 \\ 
   && Firth  & -2.181 & 0.016 & 0.319 & 0.541 & 0.944 & 2.121 \\ 
   && MPLE  & -2.227 & -0.030 & 0.305 & 0.527 & 0.966 & 2.065 \\ 
   \arrayrulecolor{black}\hline
\end{tabular}
}
\caption{Estimates, mean bias error, mean squared error, estimated standard errors, coverage probabilities, and 95\% Wald confidence intervals based on 1000 simulations with different sample sizes. The true Emax model parameters are fixed at $E_0 = -2.197$, $E_{max}=3.583$, and $log(ED_{50})=2.015$.}
\label{Sim:estimate}
\end{table}

\begin{table}[ht]
\centering
\resizebox*{!}{0.9\textwidth}{
\begin{tabular}{lll|rrrrrr}
  \hline
 Maximal Achievable Effect&  Parameter &Type  & Estimate & MBE & MSE & Est.SE & CP & Est.Length \\ 
  \arrayrulecolor{black}\hline\arrayrulecolor[gray]{0.8}
30\%($E_{max}=1.349$) & log($ED_{50}$)& MLE  & 2.438 & 0.423 & 3.980 & 1.871 & 0.991 & 7.333 \\ 
   && Cox-Snell   & 2.368 & 0.353 & 4.011 & 1.894 & 0.810 & 7.424 \\ 
   && Firth &   2.286 & 0.272 & 3.319 & 1.975 & 0.924 & 7.743 \\ 
   && MPLE   & 2.275 & 0.260 & 1.291 & 1.833 & 1.000 & 7.184 \\ 
   \hline
   & $E_{max}$& MLE  & 1.720 & 0.370 & 1.197 & 0.790 & 0.975 & 3.098 \\ 
   && Cox-Snell   & 0.977 & -0.373 & 3.210 & 0.783 & 0.789 & 3.069 \\ 
   && Firth &   1.637 & 0.287 & 1.144 & 0.707 & 0.940 & 2.770 \\ 
   && MPLE   & 1.561 & 0.211 & 0.372 & 0.689 & 0.988 & 2.701 \\ 
   \hline
   & $E_0$& MLE  & -2.312 & -0.115 & 0.514 & 0.534 & 0.965 & 2.091 \\ 
   && Cox-Snell   & -2.194 & 0.003 & 0.408 & 0.534 & 0.931 & 2.093 \\ 
   && Firth   & -2.275 & -0.078 & 0.485 & 0.534 & 0.968 & 2.093 \\ 
   && MPLE   & -2.288 & -0.090 & 0.249 & 0.540 & 0.984 & 2.118 \\ 
 \arrayrulecolor{black}\hline\arrayrulecolor[gray]{0.8}
50\%($E_{max}=2.197$) & log($ED_{50}$)& MLE  & 2.235 & 0.220 & 1.368 & 1.095 & 0.968 & 4.292 \\ 
   && Cox-Snell   & 2.313 & 0.298 & 1.103 & 1.043 & 0.948 & 4.089 \\ 
   && Firth   & 1.949 & -0.066 & 1.893 & 1.179 & 0.926 & 4.621 \\ 
   && MPLE   & 2.161 & 0.146 & 0.733 & 0.988 & 0.970 & 3.873 \\ 
   \hline
   & $E_{max}$& MLE  & 2.409 & 0.212 & 0.622 & 0.679 & 0.980 & 2.662 \\ 
   && Cox-Snell   & 2.034 & -0.163 & 1.060 & 0.659 & 0.936 & 2.583 \\ 
   && Firth   & 2.329 & 0.132 & 0.600 & 0.638 & 0.982 & 2.502 \\ 
   && MPLE   & 2.308 & 0.111 & 0.325 & 0.623 & 0.984 & 2.443 \\ 
   \hline
   & $E_0$& MLE  & -2.297 & -0.099 & 0.497 & 0.552 & 0.954 & 2.164 \\ 
   && Cox-Snell   & -2.104 & 0.093 & 0.184 & 0.557 & 0.954 & 2.183 \\ 
   && Firth &   -2.201 & -0.004 & 0.376 & 0.537 & 0.960 & 2.103 \\ 
   && MPLE  & -2.243 & -0.046 & 0.275 & 0.541 & 0.960 & 2.122 \\ 
   \arrayrulecolor{black}\hline\arrayrulecolor[gray]{0.8}
70\%($E_{max}=3.044$) & log($ED_{50}$)& MLE  & 1.949 & -0.066 & 0.389 & 0.585 & 0.984 & 2.293 \\ 
   && Cox-Snell   & 2.111 & 0.096 & 0.298 & 0.525 & 0.982 & 2.058 \\ 
   && Firth   & 1.958 & -0.057 & 0.538 & 0.615 & 0.976 & 2.411 \\ 
   & &MPLE   & 2.014 & -0.001 & 0.256 & 0.560 & 0.982 & 2.194 \\ 
   \hline
   & $E_{max}$& MLE  & 3.230 & 0.185 & 0.478 & 0.636 & 0.980 & 2.492 \\ 
   && Cox-Snell   & 2.993 & -0.051 & 0.304 & 0.663 & 0.972 & 2.599 \\ 
   && Firth &   3.138 & 0.093 & 0.418 & 0.637 & 0.976 & 2.496 \\ 
   && MPLE   & 3.132 & 0.087 & 0.314 & 0.610 & 0.980 & 2.390 \\ 
   \hline
   & $E_0$& MLE  & -2.347 & -0.150 & 0.419 & 0.577 & 0.976 & 2.262 \\ 
   && Cox-Snell   & -2.152 & 0.045 & 0.250 & 0.571 & 0.970 & 2.238 \\ 
   && Firth &   -2.228 & -0.031 & 0.360 & 0.568 & 0.968 & 2.228 \\ 
   && MPLE   & -2.271 & -0.074 & 0.253 & 0.552 & 0.970 & 2.163 \\    
  \arrayrulecolor{black}\hline
\end{tabular}
}
\caption{Estimates, mean bias error, mean squared error, estimated standard errors, coverage probabilities, and 95\% Wald confidence intervals based on 1000 simulations with different true maximal achievable effects in percentage. The sample size is fixed $n=200$, $E_0 = -2.197$, and $log(ED_{50})=2.015$.}
\label{Sim:estimate_emax}
\end{table}

\begin{table}[ht]
\centering
\resizebox{0.9\textwidth}{!}{
\begin{tabular}{rllrrrrrr}
  \arrayrulecolor{black}\hline\arrayrulecolor[gray]{0.8}
 $ED_{50}$&  Parameter &Type  & Estimate & MBE & MSE & Est.SE & CP & Est.Length \\
  \arrayrulecolor{black}\hline\arrayrulecolor[gray]{0.8}
25 & log($ED_{50}$) & MLE  & 3.386 & -0.233 & 1.339 & 1.113 & 0.966 & 4.362 \\ 
  & & Cox-Snell & 2.679 & -0.540 & 2.234 & 1.113 & 0.958 & 4.363 \\ 
  & & Firth& 3.181 & -0.038 & 1.588 & 1.109 & 0.960 & 4.347 \\ 
  & & MPLE  & 3.132 & -0.086 & 0.892 & 1.105 & 0.982 & 4.330 \\ 
  \hline
   & $E_{max}$ & MLE  & 2.623 & 0.425 & 0.766 & 0.939 & 0.974 & 3.682 \\ 
  & & Cox-Snell & 1.810 & -0.388 & 1.960 & 0.904 & 0.898 & 3.544 \\ 
  & & Firth  & 2.513 & 0.316 & 0.693 & 0.779 & 0.962 & 3.054 \\ 
  & & MPLE  & 2.407 & 0.209 & 0.366 & 0.735 & 0.986 & 2.882 \\ 
  \hline
   & $E_0$ & MLE  & -2.386 & -0.189 & 0.328 & 0.540 & 0.982 & 2.118 \\ 
  & & Cox-Snell  & -2.296 & -0.099 & 0.153 & 0.542 & 0.992 & 2.124 \\ 
  & & Firth  & -2.382 & -0.185 & 0.393 & 0.529 & 0.974 & 2.075 \\ 
  & & MPLE  & -2.350 & -0.153 & 0.251 & 0.534 & 0.984 & 2.092 \\ 
   \arrayrulecolor{black}\hline\arrayrulecolor[gray]{0.8}
50 & log($ED_{50}$) & MLE  & 3.545 & -0.367 & 2.080 & 1.560 & 0.954 & 6.114 \\ 
  & & Cox-Snell  & 2.616 & -1.296 & 4.623 & 1.548 & 0.821 & 6.068 \\ 
  & & Firth  & 3.751 & -0.161 & 3.024 & 1.539 & 0.926 & 6.033 \\ 
  & & MPLE  & 3.638 & -0.274 & 1.117 & 1.336 & 0.939 & 5.238 \\ 
  \hline
   & $E_{max}$& MLE  & 2.792 & 0.595 & 1.931 & 1.193 & 0.975 & 4.676 \\ 
  & & Cox-Snell  & 1.130 & -1.067 & 5.214 & 1.157 & 0.738 & 4.535 \\ 
  & & Firth  & 2.609 & 0.412 & 2.273 & 1.094 & 0.918 & 4.287 \\ 
  & & MPLE  & 2.362 & 0.165 & 0.503 & 0.940 & 0.979 & 3.683 \\ 
  \hline
   & $E_0$& MLE  & -2.419 & -0.222 & 0.685 & 0.547 & 0.985 & 2.144 \\ 
  & & Cox-Snell  & -2.527 & -0.330 & 0.680 & 0.562 & 0.920 & 2.203 \\ 
  & & Firth  & -2.518 & -0.321 & 0.813 & 0.587 & 0.937 & 2.302 \\ 
  & & MPLE  & -2.371 & -0.174 & 0.270 & 0.518 & 0.994 & 2.029 \\ 
   \arrayrulecolor{black}\hline\arrayrulecolor[gray]{0.8}
 75 & log($ED_{50}$) & MLE  & 3.878 & -0.439 & 2.121 & 1.770 & 0.951 & 6.940 \\ 
  & & Cox-Snell  & 1.897 & -2.421 & 11.930 & 1.632 & 0.683 & 6.397 \\ 
  & & Firth  & 3.664 & -0.654 & 4.946 & 1.611 & 0.876 & 6.315 \\ 
  & & MPLE  & 3.820 & -0.497 & 1.296 & 1.542 & 0.932 & 6.044 \\
  \hline
   & $E_{max}$& MLE  & 2.878 & 0.681 & 2.657 & 1.542 & 0.979 & 6.046 \\ 
  & & Cox-Snell  & 0.487 & -1.710 & 9.292 & 1.242 & 0.570 & 4.869 \\ 
  & & Firth  & 2.184 & -0.013 & 3.447 & 1.131 & 0.880 & 4.433 \\ 
  & & MPLE  & 2.272 & 0.075 & 0.638 & 1.070 & 0.979 & 4.195 \\
  \hline
   & $E_0$& MLE  & -2.424 & -0.227 & 0.688 & 0.495 & 0.977 & 1.942 \\ 
  & & Cox-Snell  & -2.726 & -0.529 & 1.158 & 0.533 & 0.826 & 2.089 \\ 
  & & Firth  & -2.657 & -0.460 & 1.194 & 0.617 & 0.897 & 2.420 \\ 
  & & MPLE  & -2.391 & -0.193 & 0.269 & 0.514 & 0.998 & 2.014 \\ 
   \arrayrulecolor{black}\hline\arrayrulecolor[gray]{0.8}  
\end{tabular}
}
\caption{Estimates, mean bias error, mean squared error, estimated standard errors, coverage probabilities, and 95\% Wald confidence intervals based on 1000 simulations with different true half-maximal effect dosages. The sample size is fixed $n=200$, $E_0 =  -2.197$, and $Emax = 2.197$.}
\label{Sim:estimate_ed50}
\end{table}

\begin{table}[htbp]
    \centering
    \begin{tabular}{l|ccccc}
    \hline
        Dose(mg) & Placebo & 7.5 & 22.5 & 75 & 225 \\
        \hline
         Sample size& 73 & 71 & 72 & 71 & 70\\
         Missing response & 6 &8 &1 & 3& 6\\
         Remission (Yes) & 2 & 8& 12 & 11 & 4\\
              \hline
    \end{tabular}
    \caption{Sample size, the number of missing response cases, and the number of remission cases for each dosage group in TURANDOT study.}
    \label{Table::turandot_descriptive}
\end{table}

\begin{table}[htp]
    \centering
    {\scriptsize
    \begin{tabular}{lc|ccc}
    \hline
         Parameter& Method& Estimate & StdErr & 95\% CI  \\
        \arrayrulecolor{black}\hline\arrayrulecolor[gray]{0.8}
         log($ED_{50}$)& MLE & 0.480 & 1.856 & (-3.159, 4.119)\\
                     & Cox-Snell & 3.948 & 3.345 & (-2.608, 10.504)\\
                     & Firth & 0.001 & 2.448 & (-4.798, 4.799)\\
                     & MPLE & 1.058 & 0.836 & (-0.580, 2.696)\\
        \hline
         $E_{max}$& MLE & 1.938 & 0.788 & (0.394, 3.481)\\
                & Cox-Snell & 1.746 & 2.166 & (2.499, 5.991)\\
                & Firth & 1.924 & 0.722 & (0.508,3.339)\\
                & MPLE & 1.989 & 0.718 & (0.581, 3.340)\\
        \hline
         $E_{0}$& MLE & -3.484 & 0.718 & (-4.890, -2.077)\\
              & Cox-Snell & -3.022 & 0.540 & (-4.080, -1.963)\\
              & Firth & -3.295 & 0.658 & (-4.584, -2.005)\\
              & MPLE & -3.486 & 0.640 & (-4.741, -2.232)\\
       \arrayrulecolor{black}\hline\arrayrulecolor[gray]{0.8}
    \end{tabular}
    }
     \caption{Analysis result of TURANDOT data with different bias reduction estimation methods.}
     
    \label{Table::turandot_analysis}
\end{table}

\clearpage
\input{supplmentary_arxiv}

\end{document}

%% file: supplmentary_arxiv.tex
\newpage

\setcounter{page}{1}

\begin{center}
\textit{\large Supplementary material to}
\end{center}
\begin{center}
{\large Evaluating Bias Reduction Methods in Emax Model for Reliable Dose-Response Estimation}
\vskip10pt
\end{center}

\setcounter{section}{0}
\renewcommand{\thesection}{\Alph{section}}

\maketitle
\begin{abstract}
  This document contains the supplementary material to the paper
  ``Evaluating Bias Reduction Methods in Emax Model for Reliable Dose-Response Estimation".
\end{abstract}

\section{Notations and Symbols}
We present all the notations and symbols used in Section 2-4 here as Table \ref{table::S1}.
\begin{table}[ht]
\centering
\caption{Summary of notation.}
\resizebox*{!}{0.9\textheight}{
\begin{tabular}{llp{9cm}}
\hline
\textbf{Symbol} & \textbf{Type} & \textbf{Definition} \\
\hline
$n$ & scalar & Total sample size. \\
$J$ & scalar & Number of dose levels. \\
$N$ & scalar & Number of simulation replicates.\\
$D_1,\ldots,D_J$ & set & Distinct dose levels used in the study. \\
$\text{Dose}_i$ & scalar & Dose assigned to subject $i$; $\text{Dose}_i \in \{D_1,\ldots,D_J\}$. \\
$y_i$ & scalar & Binary response for subject $i$ ($0/1$). \\
$\pi_i$ & scalar & Success probability for subject $i$: $\pi_i = P(y_i=1 \mid \text{Dose}_i)$. \\
$E_0$ & scalar & Expected baseline effect in logit at dose $0$ (placebo arm). \\
$E_{\max}$ & scalar & Expected maximal effect in logit relative to $E_0$. \\
$\mathrm{ED}_{50}$ & scalar & Dose producing half of $E_{\max}$. \\
$\lambda$ & scalar & Hill/slope parameter. \\
$\utheta$ & vector & Model parameters; here $\utheta=(E_0,\,E_{\max},\,\mathrm{ED}_{50})$. \\
$\eta(\text{Dose}_i,\utheta)$ & function & non-Linear predictor: $\eta=\mathrm{logit}(\pi_i)=E_0+\dfrac{E_{\max}\,\text{Dose}_i}{\mathrm{ED}_{50}+\text{Dose}_i}$ (for $\lambda=1$). \\
$\nabla \eta(\text{Dose}_i,\utheta)$ & vector & Gradient of $\eta$ w.r.t.\ $\utheta$. \\
$U(\utheta)$ & vector & Score function: $U(\utheta)=\sum_{i=1}^n (y_i-\pi_i)\,\nabla \eta(\text{Dose}_i,\utheta)$. \\
$U_s$ & scalar & $s$th component of $U(\utheta)$. \\
$H(\utheta)$ & matrix & Hessian of $\ell(\utheta)$ (as derived in the text); observed information is $-H(\utheta)$. \\
$I(\utheta)$ & matrix & Expected Fisher information. \\
$A_i(\utheta)$ & matrix & Additional matrix in Hessian due to non-linear model in (7). \\
$Q_i(\utheta)$ & matrix & Matrix used in Firth’s adjustment term in (11). \\
$L(\utheta\,|\,y,\text{Dose})$ & scalar & Likelihood function. \\
$L^*(\utheta)$ & scalar & Jeffreys-penalized likelihood. \\
$W(\utheta)$ & vector & additional term in Firth's score modification in (9). \\
$\tilde{U}_s(\utheta)$ & scalar & Firth's modified score component. \\
$U^*_s(\utheta)$ & scalar & Jeffrey's prior penalized score component. \\
$O^*(\utheta)$ & matrix & Observed information for penalized log-likelihood. \\
$\kappa^{rj}$ & scalar & $(r,j)$ element of the inverse of the \emph{negative} expected information (Cox–Snell expansion). \\
$\kappa_{rjl}$ & scalar & Joint cumulant: $\kappa_{rjl}=E\!\left[\dfrac{\partial^3 \ell}{\partial \utheta_r\partial \utheta_j\partial \utheta_l}\right] = E\!\left[\dfrac{\partial H_{rj}}{\partial \utheta_l}\right]$. \\
$\kappa_{rj,l}$ & scalar & Joint cumulant: $\kappa_{rj,l}=E\!\left[\left(\dfrac{\partial^2 \ell}{\partial \utheta_r\partial \utheta_j}\right)\!\left(\dfrac{\partial \ell}{\partial \utheta_l}\right)\right]=E[H_{rj}U_l]$. \\
$\kappa^{(k)}_i$ & scalar & $k$th cumulant of $y_i$. \\
$\hat\utheta$ & vector & MLE. \\
$\hat{\utheta}_C$ & vector & Cox–Snell bias-corrected estimator. \\
$\hat\utheta_F$ & vector & Firth bias-reduced estimator. \\
$\hat\utheta_J$ & vector & MPLE under Jeffreys prior. \\
$\mathrm{tr}(\cdot)$ & operator & Matrix trace. \\
$|\cdot|$ & operator & Determinant. \\
\hline
\end{tabular}
}
\label{table::S1}
\end{table}

\section{Derivation of $\kappa_{i j l}$ and $\kappa_{i j, l}$}
Since
\begin{equation*}
    \begin{split}
        \kappa_{r j l} &= \mathrm{E}\left[\frac{\partial^3 l}{\partial \utheta_r \partial \utheta_j \partial \utheta_l}\right] = \mathrm{E}\left[\frac{\partial H_{rj}}{\partial \utheta_l}\right],\\
        \kappa_{r j, l} &= \mathrm{E}\left[\left(\frac{\partial^2 l}{\partial \utheta_r \partial \utheta_j}\right)\left(\frac{\partial l}{\partial \utheta_l}\right)\right] =  \mathrm{E}\left[ H_{rj} U_{l}\right].
    \end{split}
\end{equation*}
For $\kappa_{r j l}$, we let $\kappa_{..l} =\mathrm{E}\left[\partial H/\partial \utheta_l\right]$, where $\utheta_1 =E_0$, $\utheta_2 =E_{max}$, and $\utheta_3 =ED_{50}$. Then
\begin{align*}
    \kappa_{..1} &= \mathrm{E}\left[\frac{\partial H}{\partial \utheta_1}\right]\\
&= \mathrm{E}\left[\sum_{i=1}^n \frac{\partial }{\partial \utheta_1}\left(\pi_{i}-1\right) \pi_{ij} \nabla \eta\left(\text {Dose}_{i}, \boldsymbol{\utheta}\right)^{\top} \nabla \eta\left(\text {Dose}_{i}, \boldsymbol{\utheta}\right)-\frac{\partial }{\partial \utheta_1}A_i(\boldsymbol{\theta})\right]\\
&= \sum_{i=1}^n \left(2\pi_{i}-1\right)\pi_{i}(1-\pi_{i}) \nabla \eta\left(\text {Dose}_{i}, \boldsymbol{\utheta}\right)^{\top} \nabla \eta\left(\text {Dose}_{i}, \boldsymbol{\utheta}\right)-\frac{\partial }{\partial \utheta_1}A_i(\boldsymbol{\theta})
\end{align*}
where 
$$\frac{\partial }{\partial \utheta_1}A_i(\boldsymbol{\theta}) = \left(\begin{array}{ccc}
0 & 0 & 0 \\
0 & 0 & \frac{-\pi_{i}\left(1-\pi_{i}\right) \text {Dose}_{i}}{\left(ED_{50}+\text {Dose}_{i}\right)^2} \\
0 & \frac{-\pi_{i}\left(1-\pi_{i}\right) \text {Dose}_{i}}{\left(ED_{50}+\text {Dose}_{i}\right)^2} & \frac{2\pi_{i}\left(1-\pi_{i}\right) \text {Dose}_{i} \times E_{\max}}{\left(ED_{50}+\text {Dose}_{i}\right)^3}
\end{array}\right);$$

\begin{align*}
    \kappa_{..2} =& \mathrm{E}\left[\frac{\partial H}{\partial \utheta_2}\right]\\
                 =& \mathrm{E}\left[\sum_{i=1}^n \frac{\partial }{\partial \utheta_2}\left(\pi_{i}-1\right) \pi_{ij} \nabla \eta\left(\text {Dose}_{i}, \boldsymbol{\utheta}\right)^{\top} \nabla \eta\left(\text {Dose}_{i}, \boldsymbol{\utheta}\right)-\frac{\partial }{\partial \utheta_2}A_i(\boldsymbol{\theta})\right]\\ 
                 =& \sum_{i=1}^n \left(2\pi_{i}-1\right)\pi_{i}(1-\pi_{i}) \nabla \eta\left(\text {Dose}_{i}, \boldsymbol{\utheta}\right)^{\top} \nabla \eta\left(\text {Dose}_{i}, \boldsymbol{\utheta}\right) \frac{\text {Dose}_{i}}{ED_{50}+\text {Dose}_{i}} - \\
                 & 2\pi_{i}(1-\pi_{i})\unu \nabla \eta\left(\text {Dose}_{i}, \boldsymbol{\utheta}\right) - D_i(\utheta)
\end{align*} 
where $$\unu = \left[0,0,-\frac{\text {Dose}_{i}}{(ED_{50}+\text {Dose}_{i})^2}\right],$$
and $$D_i(\utheta) = \left(\begin{array}{ccc}
0 & 0 & 0 \\
0 & 0 & \frac{-\pi_{i}\left(1-\pi_{i}\right) \text {Dose}^2_{i}}{\left(ED_{50}+\text {Dose}_{i}\right)^3} \\
0 & \frac{-\pi_{i}\left(1-\pi_{i}\right) \text {Dose}^2_{i}}{\left(ED_{50}+\text {Dose}^2_{i}\right)^3} & \frac{2\pi_{i}\left(1-\pi_{i}\right) \text {Dose}^2_{i} \times E_{\max}}{\left(ED_{50}+\text {Dose}_{i}\right)^4}
\end{array}\right);$$

\begin{align*}
    \kappa_{..3} =& \mathrm{E}\left[\frac{\partial H}{\partial \utheta_3}\right]\\
                 =& \mathrm{E}\left[\sum_{i=1}^n \frac{\partial }{\partial \utheta_3}\left(\pi_{i}-1\right) \pi_{ij} \nabla \eta\left(\text {Dose}_{i}, \boldsymbol{\utheta}\right)^{\top} \nabla \eta\left(\text {Dose}_{i}, \boldsymbol{\utheta}\right)-\frac{\partial }{\partial \utheta_3}A_i(\boldsymbol{\theta})\right]\\ 
                 =& \sum_{i=1}^n -\left(2\pi_{i}-1\right)\pi_{i}(1-\pi_{i}) \nabla \eta\left(\text {Dose}_{i}, \boldsymbol{\utheta}\right)^{\top} \nabla \eta\left(\text {Dose}_{i}, \boldsymbol{\utheta}\right) \frac{ \text {Dose}_{i} \times E_{\max}}{\left(ED_{50}+\text {Dose}_{i}\right)^2} - \\
                 & 2\pi_{i}(1-\pi_{i})\utau \nabla \eta\left(\text {Dose}_{i}, \boldsymbol{\utheta}\right) - E_i(\utheta)
\end{align*} 
where  $$\utau = \left[0,-\frac{\text {Dose}_{i}}{(ED_{50}+\text {Dose}_{i})^2},\frac{2\text {Dose}_{i} \times E_{\max}}{(ED_{50}+\text {Dose}_{i})^3}\right],$$
and $$E_i(\utheta) = \left(\begin{array}{ccc}
0 & 0 & 0 \\
0 & 0 & \frac{\pi_{i}\left(1-\pi_{i}\right) \text {Dose}^2_{i}\times E_{\max}}{\left(ED_{50}+\text {Dose}_{i}\right)^4} \\
0 & \frac{\pi_{i}\left(1-\pi_{i}\right)  \text {Dose}^2_{i}\times E_{\max}}{\left(ED_{50}+\text {Dose}_{i}\right)^4} & \frac{-2\pi_{i}\left(1-\pi_{i}\right) \text {Dose}^2_{i} \times E_{\max}^2}{\left(ED_{50}+\text {Dose}_{i}\right)^5}
\end{array}\right).$$

Then for $\kappa_{r j, l}$, we let $\kappa_{..,l} =\mathrm{E}\left[ H U_l\right]$, where $\utheta_1 =E_0$, $\utheta_2 =E_{max}$, and $\utheta_3 =ED_{50}$. Note that $\mathrm{E}\left[y_i-\pi_{i}\right] = 0$ and $\mathrm{E}\left[(y_i-\pi_{i})(y_j-\pi_{j})\right] = 0$ for $i\neq j$ due to independence, then we have

\begin{align*}
    \kappa_{..,1} & = \mathrm{E}\left[H \sum_{i=1}^n (y_i-\pi_{i})\right]\\
    & = \mathrm{E}\left[-\sum_{i=1}^n A_i(\boldsymbol{\theta}) \sum_{i=1}^n (y_i-\pi_{i})\right]\\
    & = -\sum_{i=1}^n \mathrm{E}\left[ A_i(\boldsymbol{\theta})  (y_i-\pi_{i})\right]\\
    &= -\sum_{i=1}^n \left(\begin{array}{ccc}
0 & 0 & 0 \\
0 & 0 & \frac{\pi_{i}\left(1-\pi_{i}\right) \text {Dose}_{i}}{\left(ED_{50}+\text {Dose}_{i}\right)^2} \\
0 & \frac{\pi_{i}\left(1-\pi_{i}\right) \text {Dose}_{i}}{\left(ED_{50}+\text {Dose}_{i}\right)^2} & -\frac{2\pi_{i}\left(1-\pi_{i}\right) \text {Dose}_{i} \times E_{\max}}{\left(ED_{50}+\text {Dose}_{i}\right)^3}
\end{array}\right).
\end{align*}
Similarly, $\kappa_{..,2}$ and $\kappa_{..,3}$ can be derived as:
$$\kappa_{..,2} =  \frac{\text {Dose}_{i}}{\text {Dose}_{i}+E D_{50}}\kappa_{..,1}$$
$$\kappa_{..,3} =  -\frac{\text {Dose}_{i} \times E_{\max }}{(\text {Dose}_{i}+E D_{50})^2}\kappa_{..,1}.$$

\section{Simulated illustration of a unimodal (non-monotonic) curve}
We generated $n=600$ binary outcomes at doses $d\in\{0,5,10,20,40,80\}$ with equal allocation, 
from a unimodal logit model $\eta(d)=\beta_0+\beta_1 d+\beta_2 d^2$ with $\beta_0=-2,\beta_1=0.12,$ and $\beta_3=-0.015$(peak near $d=40$), then fit (i) binary Emax model with $\lambda=1$and (ii) a quadratic-logit model. 

Across 200 replicates, the E$_{\max}$ fit was stable but often pushed $\widehat{ED}_{50}$ beyond the top dose when the true curve declined at high doses, whereas the quadratic-logit model recovered the peak and yielded smaller predictive error at $d\ge 40$. This illustrates our recommendation to report E$_{\max}$ for monotone signals but include a simple sensitivity model when diagnostics suggest non-monotonicity.

\section{Simulation: High chance of non-monotonic/convex increasing sample response curve}
\begin{figure}[!htp]
    \centering
    \includegraphics[width=0.5\linewidth]{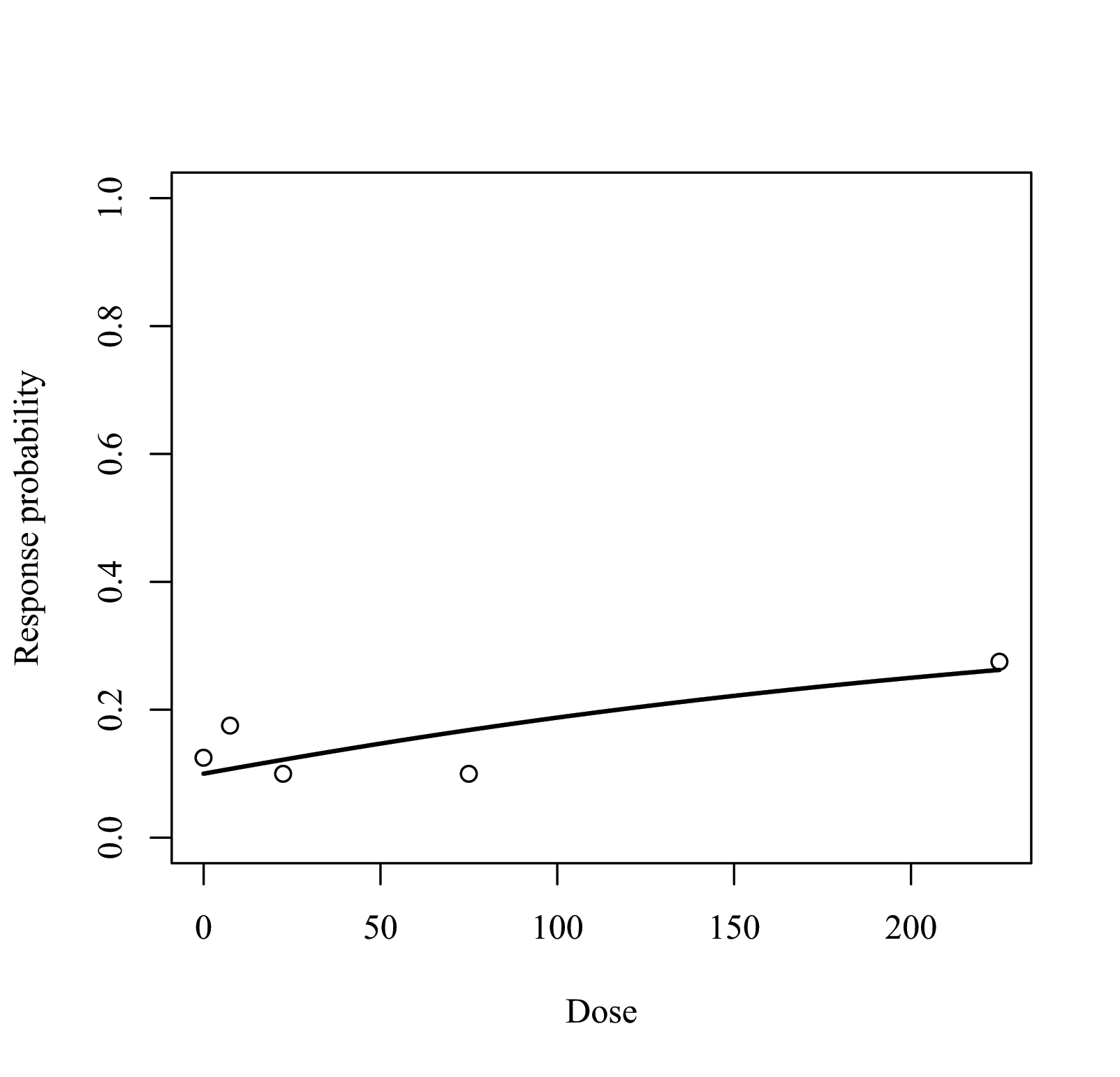}
    \caption{Demonstration of extreme pattern: Dose-response curve with $E_0=-2.197$, $E_{max}=2.197$, and $ED_{50}=250$. }
    \label{S1:convex_plot}
\end{figure}

In this section, we investigated a more extreme scenario under the Emax model, where the true $ED_{50}$ is relatively large and is not withing the dose range for the experiment. In this setting, we fixed the sample size at $n = 200$, and kept the $E_0=\mathrm{logit}(0.1)=-2.197$ and $E_{max}=\mathrm{logit}(0.5)-\mathrm{logit}(0.1)=2.197$. We changed $ED_{50}=250$ (log($ED_{50}$)=3.219). $N=1000$ replications were performed. The true response curve and the one sample dataset are plotted as \ref{S1:convex_plot}. The black line represents the true Emax model that generates the sample data, and the black dots represent the sample response probability under each dose level. Due to the flat shape of the true curve under the pre-defined dose range, the probability of observing a non-monotonic/convex increasing sample response curve is high. The sample data displayed in the figure is apparently in a non-monotonic increasing shape.

We fitted the four methods and reported the proportion of occurrences of non-convergence in log-likelihood estimation,
along with the proportion of unstable estimations due to large variance, as Table \ref{Sim:fail_proportion_convex}. It is worth noting that the porportions of failure to estimate and unstable estimates for MLE and Cox-Snell are much higher compared to the results in Section 3. Still, Firth and MPLE provide finite estimates, however, the unstable estimation proportion of the Firth method is higher.

\begin{table}[htp]
    \centering
    \begin{tabular}{l|rr}
    \hline
         Type& Fail to estimate (\%) & Unstable estimate (\%) \\
         \hline
       
          MLE & 35.6\% & 3.7\% \\
         Cox-Snell & 35.6\% & 3.7\%\\
         Firth & 0 & 11.8\%\\
         MPLE & 0& 0 \\ 
         \hline
    \end{tabular}
    \caption{Proportions of times that estimates do not exist and estimates are not stable with extreme values or variance across four methods, under high $ED_{50}$ flat increasing shape. The true Emax model parameters are fixed at $E_0 = logit(0.1)$, $E_{max}=logit(0.5)-logit(0.1)$, and $ED_{50}=250$. The sample size $N=200$.}
    \label{Sim:fail_proportion_convex}
\end{table}

Table \ref{Estimation:fail_proportion_convex} presents the estimation results for the four methods. Still, for replications where methods failed to produce an estimate, the value was recorded as NA and excluded from the metric calculations. Under this ill situation, the estimation on $ED_{50}$ for all four methods are not good, although MPLE provide the smallest MBE and MSE. Overall, MPLE consistently outperforms the other methods when evaluating MSE, yielding the smallest standard errors across all scenarios.

\begin{table}[ht]
\begin{tabular}{rllrrrrrr}
  \arrayrulecolor{black}\hline\arrayrulecolor[gray]{0.8}
 $ED_{50}$&  Parameter &Type  & Estimate & MBE & MSE & Est.SE & CP & Est.Length \\
  \arrayrulecolor{black}\hline\arrayrulecolor[gray]{0.8}
250 & log($ED_{50}$) & MLE  &3.661 & -1.637 & 5.955 & 2.394 & 0.845 & 9.386 \\ 
   &&  Cox-Snell & -3.059 & -8.358 & 71.407 & 0.861 & 0.059 & 3.373 \\ 
   && Firth & 3.821 & -1.477 & 11.615 & 2.168 & 0.748 & 8.500 \\ 
   && MPLE & 3.752 & -1.546 & 4.581 & 1.051 & 0.897 & 4.120 \\ 
   \hline
   & $E_{max}$& MLE  & 2.313 & 0.116 & 2.773 & 1.581 & 0.903 & 6.199 \\ 
   &&  Cox-Snell  & 0.105 & -2.093 & 9.369 & 0.932 & 0.466 & 3.655 \\ 
   && Firth  & 1.750 & -0.447 & 3.309 & 1.133 & 0.828 & 4.443 \\ 
   && MPLE  & 2.085 & -0.112 & 0.980 & 0.732 & 0.913 & 2.868 \\ 
   \hline
   &$E_0$& MLE  & -2.609 & -0.412 & 0.977 & 0.594 & 0.986 & 2.330 \\ 
   &&  Cox-Snell  & 1.192 & 3.389 & 19.698 & 2.184 & 0.379 & 8.560 \\ 
   && Firth  & -2.809 & -0.612 & 1.863 & 0.667 & 0.866 & 2.616 \\ 
   && MPLE  & -2.588 & -0.391 & 0.469 & 0.490 & 0.966 & 1.921 \\ 
    \arrayrulecolor{black}\hline\arrayrulecolor[gray]{0.8} 
 
\end{tabular}
   \caption{Estimates, mean bias error, mean squared error, estimated standard errors, coverage probabilities, and 95\% Wald confidence intervals based on 1000 simulations with high $ED_{50}$ flat increasing shape. The true Emax model parameters are fixed at $E_0 = -2.197$, $E_{max}=2.197$, and $log(ED_{50})=5.521$.}
   \label{Estimation:fail_proportion_convex}
\end{table}

\section{Simulation: Estimation under non-monotonic/convex increasing sample response curve}
In this section, we investigated the estimation performance under two cases where MLE tends to fail or become unstable: Case I: when the data follow a non-increasing concave shape, and Case II: when the data follow a convex increasing shape. We simulated the sample data with  $E_0 = -2.197$, $E_{max}=2.197$, and $ED_{50}=25$. The sample size, $n=210$, was evenly allocated across three different treatment dose arms: Dose=$(0, 50,150)$, as in Aletti et al.(2025). The mathematical condition for the data having an increasing concave shape is given by $\bar{y}_{D_1}<\bar{y}_{D_2}<\bar{y}_{D_3}$ and $m_1>m_2$, where

$$
m_1=\frac{\bar{y}_{D_2}-\bar{y}_{D_1}}{D_2-D_1}, \quad m_2=\frac{\bar{y}_{D_3}-\bar{y}_{D_1}}{{D_3}-{D_1}} .
$$
We checked the sampled data and categorized them into three categories: Case I: $m_1>m_2$ and $\bar{y}_{D_1}<\bar{y}_{D_2}<\bar{y}_{D_3}$ fails; Case II: $m_1<m_2$, and normal increasing concave shape. We created $N=1000$ Case I datasets and $N=1000$ Case II datasets. We fitted the four methods and present the estimation results for each method across two cases as Table \ref{Estimation:case1} and \ref{Estimation:case2}. Note for case I, both MLE and Cox-Snell methods fail to converge; thus, we only show the results for Firth and MPLE. It can be seen that in both cases, although the bias for all four methods are high when estimating $ED_{50}$, MPLE are still very stable with the lowest MSE. 

\begin{table}[ht]
\centering
\begin{tabular}{rllrrrrrr}
  \arrayrulecolor{black}\hline\arrayrulecolor[gray]{0.8}
 $ED_{50}$&  Parameter &Type  & Estimate & MBE & MSE & Est.SE & CP & Est.Length \\
  \arrayrulecolor{black}\hline\arrayrulecolor[gray]{0.8}
 25 & log($ED_{50}$)  & Score & -3.901 & -7.120 & 117.422 & 1.678 & 0.508 & 6.579 \\ 
  & & MPLE  & 2.115 & -1.103 & 1.341 & 1.294 & 1.000 & 5.071 \\ 
  &$E_{max}$ & Score  & 1.540 & -0.657 & 0.658 & 1.014 & 0.763 & 3.975 \\ 
  & & MPLE  & 1.602 & -0.595 & 0.587 & 0.505 & 0.795 & 1.978 \\ 
  &$E_0$ & Score  & -2.008 & 0.189 & 0.252 & 0.411 & 0.915 & 1.609 \\ 
  & & MPLE  & -2.095 & 0.102 & 0.217 & 0.411 & 0.949 & 1.612 \\ 
  \arrayrulecolor{black}\hline\arrayrulecolor[gray]{0.8} 
\end{tabular}
 \caption{Estimates, mean bias error, mean squared error, estimated standard errors, coverage probabilities, and 95\% Wald confidence intervals based on 1000 simulations for Case I: sample data with non-increasing concave shape. The true Emax model parameters are fixed at $E_0 = -2.197$, $E_{max}=2.197$, and $log(ED_{50})=3.218$.}
 \label{Estimation:case1}
\end{table}

\begin{table}[ht]
\centering
\begin{tabular}{rllrrrrrr}
  \arrayrulecolor{black}\hline\arrayrulecolor[gray]{0.8}
 $ED_{50}$&  Parameter &Type  & Estimate & MBE & MSE & Est.SE & CP & Est.Length \\
  \arrayrulecolor{black}\hline\arrayrulecolor[gray]{0.8}
25 & log($ED_{50}$) & MLE & 6.622 & 3.403 & 11.844 & 5.302 & 1.000 & 20.783 \\ 
  & & Cox  & -5.675 & -8.893 & 81.465 & 1.438 & 0.250 & 5.637 \\ 
  & & Score  & 7.619 & 4.400 & 23.971 & 3.158 & 0.643 & 12.381 \\ 
  & & MPLE  & 5.018 & 1.799 & 3.419 & 1.090 & 0.786 & 4.274 \\ 
  &$E_{max}$& Ori  & 10.595 & 8.398 & 116.238 & 2.885 & 1.000 & 11.308 \\
  & & Cox & -692.378 & -694.576 & $1.104\times10^6$ & 4.187 & 0.143 & 16.414 \\ 
  & & Score  & 1.386 & -0.811 & 16.061 & 1.801 & 0.643 & 7.060 \\ 
  & & MPLE  & 3.424 & 1.227 & 3.560 & 2.180 & 1.000 & 8.547 \\ 
  &$E_0$& Ori  & -1.904 & 0.294 & 0.176 & 0.412 & 0.981 & 1.615 \\ 
  & & Cox  & -99.417 & -97.220 & $1.906\times10^4$ &22.546& 0.143 & 88.380 \\ 
  & & Score  & -1.891 & 0.306 & 1.049 & 0.420 & 0.571 & 1.646 \\ 
  & & MPLE  & -2.072 & 0.125& -1.891 & 0.334 & 0.957 & 1.311 \\ 
   \arrayrulecolor{black}\hline\arrayrulecolor[gray]{0.8} 
\end{tabular}
\caption{Estimates, mean bias error, mean squared error, estimated standard errors, coverage probabilities, and 95\% Wald confidence intervals based on 1000 simulations for Case II: Sample data with convex increasing shape. The true Emax model parameters are fixed at $E_0 = -2.197$, $E_{max}=2.197$, and $log(ED_{50})=3.218$.}
\label{Estimation:case2}
\end{table}

\begin{figure}[!htp]
    \centering
    \includegraphics[width=0.5\linewidth]{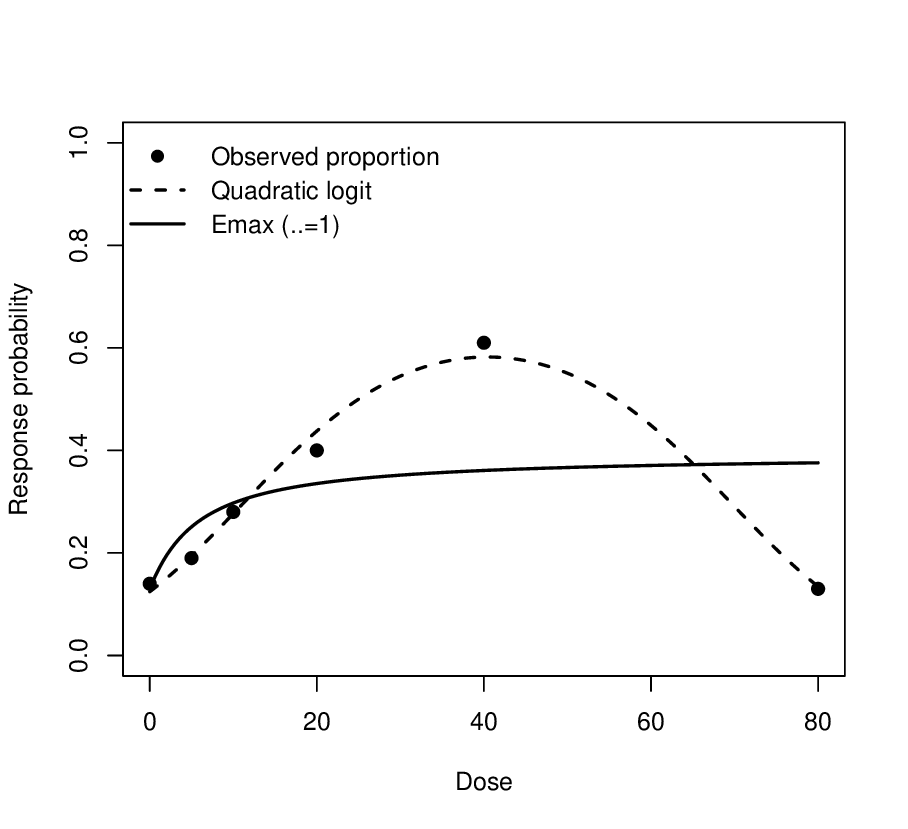}
    \caption{Demonstration: non-monotonic dose-response curve.}
    \label{S1:nonmonotonic}
\end{figure}

\newpage
